\documentclass[]{aa}
\usepackage{times}
\usepackage{graphics}
\begin{document}
\title{Interferometer $^{12}$CO observations of the box-shaped bulge spiral
 NGC\,4013}
  \subtitle{}
\author{S.Garc\'{\i}a-Burillo$^1$, F.Combes$^2$, R.Neri$^3$}
%
%
\offprints {S.Garc\'{\i}a-Burillo}
\institute {Observatorio Astronomico Nacional (IGN), 
Apartado 1143, E-28800 Alcal\'a de Henares, Madrid, SPAIN (burillo@oan.es)
\and
Observatoire de Paris, DEMIRM, 61, Av. de l'Observatoire, Paris, FRANCE 
(bottaro@obspm.fr)
\and
IRAM-Institut de Radio Astronomie Millim\'etrique, 300 Rue de la Piscine,
38406-St.Mt.d`H\`eres, FRANCE (neri@iram.fr)}

\date{Received , accepted }
%
%

\thesaurus{ 11(11.09.1 NGC\,4013; 11.09.4; 11.11.1; 11.19.2) }
\authorrunning{S.Garc\'{\i}a-Burillo et al}
\titlerunning{The box-shaped bulge spiral NGC\,4013}
\maketitle

\begin{abstract}

The nucleus of the box-shaped galaxy NGC\,4013 has been observed with the IRAM interferometer
in the J=1--0 and J=2--1 lines of $^{12}$CO. Our maps show the existence of a fast--rotating
(130 kms$^{-1}$) molecular gas disk of radius r$\sim$2$\arcsec$ (110pc). Several arguments  
support the existence of a bar potential in NGC\,4013. The {\it figure-of-eight} pattern of the 
major axis p-v plot, the ring-like distribution of gas, and the existence of gas emission at 
non-circular velocities are best accounted by a bar.

We have also detected gas at high z distances from the plane (z$\sim$200-300pc=4$\arcsec$-5$\arcsec$). 
The latter component is related to a system of 4 H$\alpha$ filaments of diffuse ionized gas 
that come out from the nucleus. The galactic fountain model seems the best to 
account for the H$\alpha$ and CO filaments. Although the peanut distortion can be spontaneously 
formed by a stellar bar in the disk, gas at high z might have been ejected after a nuclear starburst. 
The H$\alpha$ filaments start in the plane of the disk at r$\sim$200pc(4$\arcsec$), and reach several Kpc 
height at r$\sim$600pc(10$\arcsec$), coinciding with the maximum peanut distortion where the strength of the
restoring forces of the plane have a minimum. We have critically examined other 
alternatives judged less probable: 
the existence of a CO warp (connected to the HI warping disk), the accretion of gas along 
stable inclined orbits and, finally, a vertical gas 
response near the resonances of the peanut (the latter is tested by numerical simulations).   
  
Although a link between the bar and the box--shaped bulge in NGC\,4013 is suggested 
 we find noticeable differences
between the results of previous numerical simulations and the present observations. The 
discrepancy concerns the parameters of the bar generating the peanut. We see in NGC\,4013 the 
existence of a {\it strong} ILR region. The inclusion of a dissipative component, 
which remains to be thoroughly studied, may change the evolution of the stellar peanut: 
although in simulations the peanut appears initially near a {\it marginal} ILR, 
the inflow of gas driven by the bar, can make two ILRs appear.

\end{abstract} 


\section{Introduction}

N-body numerical simulations  have established a link between the existence of boxy-shaped 
galaxy bulges and the presence of a bar instability in the disk (Combes and Sanders, 1981; 
Combes et al 1990, hereafter called {\bf C90}; Raha et al 1991; Pfenniger and Friedli, 1991; 
Friedli and Martinet, 1993; Friedli and Benz, 1993).
The bar can thicken, driven by vertical resonances, giving the bulge a boxy-shaped or peanut 
appearance, provided that the galaxy is seen at a high inclination (i$>$70$^{\circ}$), and 
depending on the orientation of the major axis of the bar with respect to the line of nodes 
(measured by the azimuth $\Phi$): an edge-on bar ($\Phi$=90$^{\circ}$) adopts a boxy 
shape and the peanut feature appears for a large range of intermediate azimuths. The thickening is 
supported by star orbits that leave the galaxy plane and it occurs only in 
a {\it privileged} region along the bar where there is coincidence between the horizontal 
resonance ILR ($\Omega_{b}$\\=$\Omega$-$\kappa$/2) and its vertical counterpart 
($\Omega_{b}$=$\Omega$-$\nu_{z}$/2). Other me\-cha\-nisms, not related to the presence of a bar
potential in the disk, can explain the
formation of peanut bulges. Several authors support the merger/accretion interpretation 
(Binney and Petrou 1985, Shaw 1993, Shaw et al 1993). The occurrence of dissipationless collapse
of an isolated system 
(Lima-Neto and Combes 1995), possibly in the presence of a bound com\-pa\-nion (May et al (1985))
have also accounted for the existence of boxy shapes in galaxy bulges.

 In view of the large variety of scenarios it is necessary  
to envisage a detailed comparison between the different models and the observations 
for specific objects. There are very few examples where the connection between bars and box-peanut bulges 
have been investigated from an observational point of view. Bettoni and Galletta 1994 found photometric 
evidence of a bar in the peanut spiral NGC\,4442; Kuijken and Merrifield 1995 studied 
a sample of edge-on spirals using H$\alpha$ and optical lines as tracers of gas kinematics. 
However extinction effects can be severe in highly inclined disks where the column densities 
of gas and the associated dust can reach very high values. The use of {\it macroscopically} 
optically thin tracers such as CO can help us to get a more clear picture of the gas kinematics, 
especially in the central regions of these spirals. In particular, in this work we 
investigate the link between bars and box-peanut bulges using CO as a fair tracer of the 
molecular gas kinematics in the nuclear disk of a candidate galaxy: NGC\,4013.

The Sbc edge-on spiral NGC\,4013 has been identified as an extreme box-peanut bulge object,
 as seen in the optical pictures of van der Kruit and Searle (1982) and Jarvis (1986). 
Rand (1996) (hereafter referred as {\bf R96}) have taken deep H$\alpha$ images of 
NGC\,4013 searching for extraplanar diffuse ionized gas (DIG), discovering an impressive 
set of filaments coming out the plane in the nuclear region. He suggests this is the signature 
of a nuclear outflow, after a starburst episode related with the distorted morphology of the nucleus. 
NGC\,4013 was mapped in the J=2--1 and 1--0 lines of $^{12}$CO with the IRAM 30m telescope
(HPBW 13$\arcsec$ and 21$\arcsec$, respectively) by G\'omez de Castro and
 Garc\'{\i}a-Burillo, 1997 (hereafter referred as {\bf GCGB97}). The CO disk extends 
to r=100$\arcsec$ and it consists of a ring-like source (of radius r=30-40$\arcsec$) and 
an unresolved fast-rotating nuclear disk. The high-velocity CO component has no 
HI counterpart (see Bottema, 1987, 1995 and 1996) and it is best explained by {\bf GCGB97} 
invoking the presence  of a non-axisymmetric potential. 
Still the spatial resolution of the observations was insufficient to undertake a detailed 
analysis of gas kinematics in the nucleus of the galaxy. Furthermore the single-dish
data suggested also the existence of {\it out-of-the-plane} gas that might be connected 
either with the vertical structure of the box-peanut bulge stable orbits, or the occurrence 
of an outflow and local ejections of material, as the DIG measurements of 
{\bf R96} seem to indicate. High-resolution ($\sim$3$\arcsec$) interferometer CO data 
of NGC\,4013 are presented in this work, intending to study the distribution and kinematics 
of molecular gas in the nucleus of this galaxy and hopefully clarify the nature of the link 
between bars and box/peanut bulges. 

The study of dynamics of the gas in a peanut potential 
will deserve  detailed numerical simulations involving both the stars and the gas, to be fully 
presented in a forthcoming publication (paper II). In this paper we limited ourselves to analyse 
whether there is a spontaneous and stable vertical response of the gas to a peanut potential, using a first
run of numerical simulations.

\section{Observations}     

The $^{12}$CO(1--0) observations were made with the IRAM 
interferometer of Plateau de Bure (see description in Guilloteau et al 1992), 
between December 1995 and March 1996. During less than one third of the 
observing period we observed simultaneously the 2--1 line of $^{12}$CO.
We used the compact (CD) 4-antenna configurations, consisting of 18 baselines
ranging from 24 to 180 m. The declination of the source 
($\delta\sim$44$^\circ$) allowed good UV coverage and made possible to 
synthesize an ellipsoidal beam of FWHM=3.4$\arcsec\times$3.3$\arcsec$
in the 1--0 line, adopting natural weighting and no taper on the 
visibilities (FWHM=1.2$\arcsec$\\$\times$1$\arcsec$ in the 2--1 line). 
The $^{12}$CO(1--0) antenna half-power primary beam was 43$''$ (22$''$ in the
$^{12}$CO(2--1)). The primary beam field was centered at 
$\alpha$(J2000)=11$^{h}$58$^{m}$31.7$^{s}$, 
$\delta$(J2000)=43$^\circ$56$\arcmin$\\48.1$\arcsec$. Assuming a distance of 
11.6 Mpc (Bottema 1995), the 43$''$ primary beam corresponds to 2.4 kpc (1$\arcsec$=56pc).

A total bandwidth of 430 MHz (1120 and 560 kms$^{-1}$ for the 1--0 and 2--1 lines
respectively), centered at v=839kms$^{-1}$
 (LSR) and largely covering the range of velocities observed in
NGC\,4013 ($\pm$250kms$^{-1}$, according to {\bf GCGB97}), 
was observed with a
resolution of 2.5 and 1.25 MHz (6.5 kms$^{-1}$ and 3.3 kms$^{-1}$ at 115 and 230 GHz, 
respectively). The central 140 MHz of this band
was also observed at 115 GHz with a resolution of 0.63 MHz (1.6 kms$^{-1}$).

Data calibration was made in the standard way using the CLIC
software package (Lucas 1992). The correlator was calibrated
every 20 min with a noise diode, and the RF passband once at the beginning of each
observing run on 3C273. The relative phase of the antennas was checked every 20 min
on the nearby quasars 1308+256 and 1156+295.  The rms atmospheric phase
fluctuations were typically between 10$^\circ$ and 25$^\circ$ at 115GHz. 

The data were then cleaned to yield
channel maps. The rms noise in a 2.5 MHz-wide channel is
2.3mJy ($\Delta T_{B}$=0.018K) per 3.35$\arcsec^{2}$ beam at 115GHz (the 
1-$\sigma$ noise is of 20mJy per beam in the 2--1 line or equivalently
0.35K). 


%

\begin{figure*}
\vspace{5mm}
\caption{$^{12}$CO(1--0) velocity-channel maps observed with the Plateau de
IRAM interferometer with a resolution (HPBW) of 3.4$\arcsec\times$3.1$\arcsec$. 
Coordinates are with respect of the dynamical center (indicated by the cross).
Velocity-channels range from v=150kms$^{-1}$ to v=-183kms$^{-1}$ by steps of 6.5kms$^{-1}$.
Contour levels are -12, 12, 24, 40, 60, 80, 100 to 200mJybeam$^{-1}$ by 
steps of 30mJybeam$^{-1}$.}
\end{figure*}

\section{The CO maps}

In the following, we will denote by $x$ and $z$ the kinematical major and minor
axis of NGC\,4013 ($x$ increasing to the NE and $z$
to the NW).  All coordinates will be referred to the dynamical center (x$_C$,z$_C$)
which has been determined as follows: we first fit the position angle of the galaxy (PA) 
together with the z$_C$ position of the disk, applying a standard 
least-squares method to the integrated intensity map of fig2a. 
The inferred value for the orientation of the galaxy plane is 
$PA=66^\circ\pm$5$^\circ$, in agreement with the value derived from optical imaging. 
Finally both x$_C$ and the systemic velocity $v_{sys}$ are
 derived imposing the maximum symmetry on the the inner 10$\arcsec$ region of major 
axis p--v plot taken at z=z$_C$. We calculate (x$_C$,z$_C$)=(0,0)= 
$\alpha(J2000.0)= 11^{h}58^{m}31^{s}$.36, $\delta(J2000.0)=43^{o}56^{'}50^{''}$.9. 
This position coincides within 1$\arcsec$ with the radio-continuum sources detected at 
6cm and 21cm (note that Bottema (1995) reports a wrong position for the radio-continuum 
source) as well as with the optical center determined by Palumbo et al 1988. Similarly, 
the velocities ($v$) will be relative to $v_{sys}$=840$\pm$\\10 kms$^{-1}$ (LSR).

Fig.1 shows the $^{12}$CO(1--0) velocity-channel maps observed with the 
interferometer (oriented along x and z axes). Emission appears from v=143 to 
--170 kms$^{-1}$, concentrated in a rotating edge-on disk whose vertical structure is 
marginally resolved with our 3.4$\arcsec$ beam.  
However we notice 
'out of the plane' gas excursions mostly in the bottom-left quadrant (within the range 
$\Delta$x=(--5$\arcsec$,25$\arcsec$) and $\Delta$z=(--3$\arcsec$,--8$\arcsec$)), visible at 
certain velocities (within the range $\Delta$v=(--20kms$^{-1}$,--105kms$^{-1}$)).

At the derived dynamical center offset (0,0), CO emission displays a large spread of 
velocities: $\pm$130kms$^{-1}$. As shown in Fig.2a, which represents the $^{12}$CO(1--0) 
velocity-integrated temperature map, the observed high-velocity gas at the center is linked 
with the presence of a conspicuous CO nuclear disk in the inner 100\,pc(2$\arcsec$). The nuclear disk 
shows an east-west asymmetry: the maximum of emission is located at x$\sim$2$\arcsec$, i.e., 
eastwards with respect to the dynamical center. The remarkable asymmetry of the inner 100\,pc of 
NGC\,4013 is more clearly shown by the higher resolution 2--1 data: Fig.2b shows the 
velocity-integrated map in the 2--1 line which allows to resolve the nuclear disk radially. 
The latter displays an asymmetrical ring-like structure of radius r$\sim$1.7$\arcsec$ (95\,pc).

\begin{figure}
\vspace{5mm}
 \caption{{\bf a(top)}:$^{12}$CO(1--0) integrated intensity contours
observed with the IRAM interferometer towards the center of NGC\,4013. $x$ and
$z$ are offsets (in arcsec) with respect to the dynamical center {\bf C}.
Contours are 1 and 2 to 24Jy.kms$^{-1}$beam$^{-1}$, by steps of 
2Jykms$^{-1}$beam$^{-1}$. A zoomed view of figure {\bf a} is displayed below. 
  {\bf b(bottom)}: same as a) but for the 2--1 line of
$^{12}$CO. Contours are 4 to 15 by 1Jykms$^{-1}$beam$^{-1}$.}
\end{figure}

The overall distribution of molecular gas in the disk shows a similar 
E-W asymmetry with respect to (0,0): CO is stronger and it is more 
extended in the eastern side, up to x=35$\arcsec$, than in the western side of the 
disk, up to x=-25$\arcsec$ (the single-dish data of {\bf GCGB97} 
already showed this asymmetry holds for radii 40$\arcsec<$r$<$100$\arcsec$). The same 
asymmetry is echoed by other star formation tracers (H$\alpha$ and non-thermal radiocontinuum).
A similar asymmetry in the CO distribution is present in the inner nucleus of our 
Galaxy (Bally et al 1987, 1988) and other spirals 
(see the case of NGC\,891: Garc\'{\i}a-Burillo and Gu\'elin, 1995).

\subsection{The mass of molecular gas} 
 
If we take a CO to H$_2$ conversion factor of
X=N(H$_2$)/I$_{CO}$=2.3$\times$\\10$^{20}$cm$^{-2}$K$^{-1}$km$^{-1}$s\,(Strong et al 1988),
the H$_{2}$ mass derived from the $^{12}$CO(1--0) interferometer map is
M(H$_2$)=3$\times$10$^{8}$M$_{\sun}$. For this we assumed the distance to be D=11.6 Mpc and 
integrated the CO flux within a rectangular area of dimensions 
70$\arcsec \times$18$\arcsec$, centered on the position (0,0). 
Including the mass of Helium, the total molecular gas mass in the Bure 
field is M$_{gas}$=M(H$_2$+He)=4$\times$10$^{8}$M$_{\sun}$. The nuclear disk mass content
is derived to be close to M$_{gas}\sim$0.6$\times$10$^{8}$M$_{\sun}$ 
(this is an upper limit as we integrated I$_{CO}$ within r=2$\arcsec$ for all velocities,
including the contribution of gas seen in projection close to (0,0) but located far from
the nucleus in the plane of the galaxy).

Taking into account that the shortest spacing measured by the interferometer is $\sim$20m,
we expect to filter scales $<$L$>\sim$20-25$\arcsec$ at 115GHz
(equivalently, $<$L$>\sim$10-15$\arcsec$ at 230GHz). We have derived the fraction of the 
single-dish $^{12}$CO(1--0) and (2--1) fluxes included in the Bure maps. The zero-spacing flux 
filtered out by the interferometer is kept very low in the 1--0 line: we detect 
nearly $\sim$100$\%$ of the 30m-flux within the interferometer primary beam. On the contrary, 
the filtering is severe at 230GHz: we estimate that $\sim$1/2 of the 30m flux is missing 
in the 2--1 primary beam.

\begin{figure}
\vspace{5mm}
 \caption{{\bf a(top panel)}We overlay the $H\alpha$ extraplanar emission image of 
{\bf R96} showing the existence of 4 DIG filaments (grey scale) with the map of 
CO emission integrated within the velocity range
$\Delta$v=(0kms$^{-1}$,\,--80kms$^{-1}$)(contours). We show an overlay of the $^{12}$CO(1--0) 
velocity--channel 
maps at v=$-26$kms$^{-1}$ ({\bf b(top)}),v=$-52$kms$^{-1}$ ({\bf c(middle)})
and v=$-104$kms$^{-1}$ ({\bf d(bottom)})(contours) with a K band image of the
nucleus of NGC4013 taken from Shaw 1993 (gray scale contours).
A sketch of the 4 extraplanar DIG H$\alpha$ filaments is depicted in figures a--d (dashed lines).}
\end{figure}

%

\section{The vertical distribution of molecular gas} 

Assuming the vertical distribution to be gaussian-like
we have derived the thickness of the CO disk ($\Delta z$) deconvolving the measured FWHM 
on the $^{12}$CO(1--0) velocity-integrated intensity map of Fig.2a by our 
3.4$\arcsec\times$3.3$\arcsec$ synthesized beam. The inferred $\Delta z$ shows no systematic
radial trend along the major axis and it varies between
 1.5$\arcsec$ and 2.5$\arcsec$, i.e. translated into spatial scales $\Delta z$=80-130pc. These
 values are comparable with the thickness of the thin molecular gas disk in 
the Galaxy: 60--100pc (Bronfman et al 1988).

The existence of CO emission at high $z$ is clearly visible in the bottom left 
quadrant of Fig.2a (see also Fig.3a). Note that the lowest contour corresponds to 5$\times\sigma_{A}$, 
where a value of $\sigma_{A}$=0.2Jybeam$^{-1}$kms$^{-1}$, has been obtained through 
the expression $\sigma_{A}$=$\int_{\Delta\,v}\sigma\,d$v, taking $\sigma$=0.002Jybeam$^{-1}$ and 
$\Delta$v=100\\kms$^{-1}$. The mass of molecular gas at high $z$ is 
M(H$_2$)$\sim$1.5$\times$10$^{7}$\\M$_{\sun}$, ($\sim$5--10$\%$ of the total emission). 
The latter is a conservative lower limit as the CO-to-H$_2$ conversion factor might 
be significantly higher. The existence of vertical extensions of molecular gas in the nucleus 
were already suggested by {\bf GCGB97} using coarse resolution single-dish data.

The connection between the DIG filaments discovered by {\bf R96} and the above reported 
CO extraplanar emission is best illustrated at certain velocities.    
Figs.3b-c-d represent an overlay of a K-band image of NGC$\,$4013, showing the peanut bulge 
distortion of the nucleus in the region ($\Delta x$, $\Delta z$)=
($\pm$15$\arcsec$, $\pm$10$\arcsec$)  with the CO-velocity channel maps at v=--26,--52 
and --104kms$^{-1}$ (Figs.3b, 3c and 3d, respectively). A sketch of the DIG filaments 
of {\bf R96} is included for comparison.
{\bf R96} interpreted the impressive set of 4 DIG filaments above and below the nucleus of 
NGC$\,$4013 as the signature of a massive nuclear starburst. Supernova
explosions might inject a huge amount of energy to the gas finally lifted to high z.  
At close sight, filament A is associated with CO emission breaking out of the plane (v=--26kms$^{-1}$), 
although extraplanar gas appears to be more spread above the plane than the DIG feature (see 
v=--52kms$^{-1}$ and Fig 3a). Filament B is also associated with extraplanar CO (v=--104kms$^{-1}$).

We have estimated the gravitational potential energy of the extraplanar molecular gas 
associated with filament A (the most impressive), using the inferred positions
of CO emission above the mid-plane. 
Taking the two isothermal stellar components model of Jacobi and Kegel (1994), fitted on the luminosity 
distribution out of the plane in NGC\,4013, the resultant gravitational energy per unit mass 
$\Phi$ is

\begin{displaymath}
\Phi=\sigma_{t}^{2}\rm{ln}\,\rm{cosh}(z/h)
\end{displaymath}
\begin{displaymath}
\sigma_{t}^{2}=(\rho_{1}\sigma_{1}^{2}+\rho_{2}\sigma_{2}^{2})/(\rho_{1}+\rho_{2})
\end{displaymath}

where $\rho_{i}$ and $\sigma_{i}$ stand for the stellar mid-plane densities and velocity 
dispersions of the i=1,2 components and h is the fitted scale height (h=1.72kpc and $\sigma_t$=58.3kms$^{-1}$ 
for NGC\,4013, according to Jacobi and Kegel (1994)). The average height of the CO extraplanar emission 
observed within the interval v=--13,--78kms$^{-1}$ is z$\sim$270pc, which gives a total 
potential energy of 50$\times$10$^{51}$ergs for $M_{gas}$=4$\times$10$^{6}$M$_{\sun}$. 
We emphasize that the above equation allows to calculate the total potential energy of the 
material at height z, and therefore should be taken as a lower limit to the total input 
energy that must be given to the molecular gas in the plane to reach the observed values: 
first we have not considered 
the kinetic energy of these gaseous structures at present and we made no assumptions on the 
efficiency of the mechanism which is responsible for injecting the energy to the gas in the 
plane. Whatever the nature of this energy source is, at least 50 type II supernovae are 
required (assuming that a type II supernova releases $\sim$10$^{51}$ ergs to the ISM). This lower 
limit might transform into an order of magnitude higher number of required SN, once the radiative
and dissipative energy losses of the process are taken into account, henceforth suggesting that 
the nucleus of NGC\, 4013 might be experiencing a starburst episode. 
On the other hand, the restoring force of the galaxy plane is diminished near the position of the
maximum distortion of the potential and this limit can be slightly lowered.

\begin{figure*}
\vspace{5mm}
\caption{{\bf a(left)}:The $^{12}$CO(1--0) position-velocity diagram along the 
kinematical major axis, oriented along PA=66$^\circ$. Gray contours correspond 
to 0.010, 0.020, 0.035 to 0.200Jybeam$^{-1}$ by steps of 0.02Jybeam$^{-1}$. 
A zoomed view on the inner r=5$\arcsec$ region, where there is a 
CO nuclear disk showing high-velocity gas, is displayed on the right.     
{\bf b(right)}: same as a) but now for the 2--1 line of $^{12}$CO with a 
$\sim$1$\arcsec$ resolution in gray scale contours, linearly scaled from
 0.035 to 0.200Jybeam$^{-1}$ by steps of 0.025Jybeam$^{-1}$. Dashed contours 
stand for the same $^{12}$CO(1--0) position-velocity diagram shown in a). Notation of 
kinematical components ({\bf C},{\bf R} and {\bf F}) are explained in text.}
\end{figure*}

%

The FIR indices in NGC\,4013 do not indicate a massive starburst at the global scale of the disk: 
neither the normalised infrared fluxes, log(FIR/a$_{gal}^{2}$=-13.8 (where a$_{gal}$ is the galaxy 
major axis in arcminutes) and log(FIR/L$_B$)=0.3, nor the IRAS colors (S$_{25}$/S$_{12}$=1.5; 
S$_{100}$/S$_{25}$=30) are indicative of typical starbursts (see Huang et al, 1996 
for discussion of a large galaxy sample). However there is evidence that supports a 
starburst event in the nucleus: first, a huge fraction ($\sim$30$\%$) of the total 
radio-continuum flux at 21cm is emitted by the nuclear disk; this fraction is the highest among 
the sample of galaxies published by Hummel et al 1991. 
Most noticeably, there is an association between the H-shape of the DIG filaments and 
extraplanar CO emission with the box-peanut appearance of the bulge. A similar set of 
filaments has been discovered by Rand in a galaxy also classified as a box-peanut: NGC\,3079. 
Connection between the nuclear starburst, the distorted 
stellar potential and the CO kinematics is discussed below (sections 5 and 6).    

HI observations (Bottema 1995) showed the existence of a highly warped disk in NGC4013. 
CO emission at high z present in our observations could come from regions of the outer warped 
disk seen in projection near the nucleus. However a comparison of velocity channel maps of atomic and 
molecular gas leads us to reject this explanation: whereas CO emission at high z appears at 
negative velocities in the bottom left quadrant (Figs. 2-3), the HI warp shows up in the top 
left quadrant (NE in Fig. 2 of Bottema 1995) within the same velocity range. 
  
\section{Kinematics of molecular gas}

Fig.4a shows the $^{12}$CO(1--0) position-velocity diagram taken along the major axis.
The high-inclination of the galaxy allows us to get the whole range of radial velocities 
in the plane. Molecular gas emission is spread in a romboid-like region where we distinguish 
three velocity components: 

\begin{itemize}

\item

[I]$\,$The most outstanding component displays a large spread 
of velocities (v=$\pm$130kms$^{-1}$) towards the dynamical center 
(referred as the straight ridge {\bf C}) and it has no HI counterpart. 
The high-velocity gas corresponds to the central CO nuclear disk identified in Figs.2a-b 
and it extends from x=--3$\arcsec$ to x=3$\arcsec$. It is fully resolved in the 
2--1 line where it extends from x=--2$\arcsec$ to 2$\arcsec$; 
note that $dv/dr$ reaches a maximum value of $\sim$1000kms$^{-1}$kpc$^{-1}$
in the inner 100 pc(2$\arcsec$). 

\item

[II]$\,$Part of the 
emission is concentrated in a straight ridge ({\bf R}) which slowly drifts in 
velocity when we move along the major axis: it extends from 
(x=30$\arcsec$,\,v=--80kms$^{-1}$) to 
(x=--30$\arcsec$,\,v=100kms$^{-1}$). HI emission is detected in this region of the 
p-v plot according to Bottema 1987, 1995 and 1996.
CO emission fills unevenly the p--v space 
between {\bf C} and {\bf R}. 

\item

[III]$\,$We detect gas at non-circular velocities (or velocities 
{\it forbidden} by circular rotation) in the inner $\pm$5$\arcsec$. 
This region ({\bf F}) extends symmetrically on both sides of the 
nucleus:at x$\sim$--3.5$\arcsec$, we measure CO emission up to v$\sim$--60kms$^{-1}$, 
whereas circular rotation would impose v$>$0 for x$<$0. 
The same applies for the offset x$\sim$3.5$\arcsec$ where v$\sim$60kms$^{-1}$. 

\end{itemize}

\subsection{The major axis p-v profile}

We have derived the CO-based rotation curve (v$_{rot}^{CO}$) from the terminal 
velocities method, applied to the p--v major axis diagram. We have used the 2--1 data in the inner 
x=$\pm$2$\arcsec$, the 1--0 data from 2$\arcsec<$x$<$25$\arcsec$, and finally added the 30m data 
of {\bf GCGB97} for the outer disk (25$\arcsec<$x$<$100$\arcsec$). 
In the derivation, we have taken into account the effect of channel smearing 
($\Delta$v=6.5kms$^{-1}$) and assumed a typical cloud-cloud velocity dispersion of 
$\sim$10kms$^{-1}$.  Finally we have assumed that $v$ is translatable 
into v$_{rot}^{CO}$, as circular-motions should be the main contributor to $v$ (Sofue 1996; 
Garc\'{\i}a-Burillo 1997). 

Note that the HI-based rotation curve (v$_{rot}^{HI}$) cannot account for the CO high-velocity gas 
component {\bf C}. The scarcity of atomic hydrogen in the nucleus, together with the 
low resolution of HI observations can explain the differences between v$_{rot}^{HI}$ and 
v$_{rot}^{CO}$. We therefore conclude that the real rotation curve is certainly much steeper in the 
inner 20$\arcsec$ than inferred using HI data: v$_{rot}^{CO}$=110kms$^{-1}$ at r=110pc,
this implies a dynamical mass of M$_{dyn}$=r$\times v_{rot}^{2}/G$=
2.8$\times$10$^{8}$M$_{\sun}$, assuming a spheroidal component in the nuclear region 
(a factor 0.6 lower in the case of a flat disk distribution). 
Therefore, we estimate M$_{gas}$/M$_{dyn}$ to be of 22$\%$ inside the nuclear 
disk. The latter ratio goes down monotonically to reach 10$\%$ at r=500pc.      
Although the dynamics is still dominated by the stars, the contribution of molecular gas 
to the total mass content of the nuclear disk is significantly high and the effects
of gas self-gravitation might not be negligible.

As stated above, the {\bf F} component indicates the existence of emission at non-circular velocities. 
The high-resolution of the present observations precludes any 
explanation of {\bf F} in terms of beam dilution: at x=$\pm$3.5$\arcsec$ we are off by more than 
one synthesized beam from the center.

\begin{figure}
\vspace{5mm}
\caption{The radial variation of main frequencies of NGC\,4013 
($\Omega$, $\Omega$-$\kappa$/2, $\Omega$+$\kappa$/2 and  $\Omega$+$\kappa$/4) is plotted, assuming the 
epicyclic approximation and based on the CO rotation curve. The loci of the 
principal resonances are determined for a value of $\Omega_p$=65kms$^{-1}$kpc$^{-1}$.}
\end{figure}

\begin{figure}
\vspace{5mm}
\caption{The $^{12}$CO(2--1) isovelocity contours (solid and dashed lines) in the nuclear disk of
NGC\,4013, linearly spaced from -100kms$^{-1}$ to 100kms$^{-1}$ by steps of 20kms$^{-1}$, overlaid on the
corresponding integrated intensity map (fig.2b).}
\end{figure}

%
%
%

\subsection{A bar in NGC\,4013?}

The presence of the {\bf F} component is more readily explained by invoking a deviation from 
axisymmetry in the inner (r$\sim$200pc) mass distribution of NGC\,4013. More precisely, we 
have observational evidence that the gas flow in this galaxy might be driven by a 
barred potential:

\begin{itemize}

\item

The major axis p-v diagram shown in Fig.4a displays the {\it figure-of-eight} shape typical of 
gas flowing along a bar (Binney et al 1991; 
Garc\'{\i}a-Burillo and Gu\'elin 1995; Kuijken and Merrifield 1995). 
Components {\bf C} and {\bf R}, producing a 
double-peaked line-of-sight velocity plot, would correspond to molecular gas populating 
two different families of orbits: x$_2$ inner orbits for {\bf C}, and higher energy 
x$_1$ orbits for {\bf R}. The {\bf F} component is best explained as the projection of the
outer x$_1$ orbits close to the cusped orbit.  x$_2$ orbits develop between two inner Linblad 
resonances ILRs (the outer (oILR) and the inner (iILR)). x$_1$ orbits exist between the oILR 
and the corotation of the bar (COR). This particular morphology of the p-v plot cannot be 
reproduced if we impose axisymmetry in the potential (Kuijken and Merrifield 1995).

\item

Moreover we expect to see the imprint of a bar in the radial distribution of molecular gas 
as a reflect of secular evolution. Gravitational torques induce a radial redistribution of 
gas which accumulates in rings between the ILRs, the 4:1 or UHR resonance and the outer Linblad 
resonance (OLR). As a result, the corotation region is progressively emptied of gas. The radial 
distribution of gas in NGC\,4013 shows a compact central source of radius r$\sim$2$\arcsec$ ({\bf C}), 
a region relatively emptied of gas between {\bf C} and {\bf R} (r$\sim$10-30$\arcsec$) and a 
secondary maximum towards r$\sim$50$\arcsec$ (according to the single-dish data of {\bf GCGB97}). 
This distribution, showing the existence of several rings, is well accounted by the bar 
hypothesis.       

\item

Fig.6 shows the $^{12}$CO(2--1) isovelocity contours in the nuclear disk region. 
The z-distribution of molecular gas is spatially resolved at 1$\arcsec$ resolution: we detect 
the presence of a slight velocity gradient along the minor axis of the galaxy and a westwards tilt 
of $\sim$60$^{\circ}$ in the isovelocities of the nuclear disk, suggesting that we are seeing in 
projection a non-axisymmetric gas distribution.  
   
\end{itemize}  
	
Although the arguments enumerated above cannot be taken as a proof of the existence of
a barred potential, it all points out to this scenario as the simplest explanation. To
explore on more firm grounds the consequences of this hypothesis, in the following section 
we try to locate the main resonances of the bar in the disk.

\subsection{Inferring limits on the bar pattern speed}

The standard first-order approach consists of deriving the principal frequencies of the disk from the 
fitted rotation curve (v$_{rot}^{CO}$): $\Omega$, $\Omega$-$\kappa$/2, $\Omega$+$\kappa$/2 and 
$\Omega$+$\kappa$/4 (see Fig.5). Assuming the epicyclic approximation, we can figure out the value of 
the bar pattern speed ($\Omega_p$) and consequently the position of the main resonances, based on 
observational and theoretical arguments. This procedure is not intended to provide an accurate 
fit of $\Omega_p$, but to test if the basic results of numerical simulations are supported by the present
observations.

As shown in Fig.5, the $\Omega$-$\kappa$/2 curve presents a strong maximum 
($\sim$180kms$^{-1}$kpc$^{-1}$) at r$\sim$3.5$\arcsec$ and it goes monotonically down to 1/10  of 
its peak value ($\sim$20kms$^{-1}$kpc$^{-1}$) at r$\sim$20$\arcsec$; farther out, it stays quite flat 
until the edge of the optical disk.  
Self-consistent numerical simulations of bars, including only the stellar component ({\bf C90}), predict that 
the formation of the peanut occurs when the bar breaks up the plane near a {\it marginal}
 ILR (i.e., the bar pattern speed remains always close to the maximum of $\Omega$-$\kappa$/2).
If this is to apply in NGC\,4013, we would require the bar to have an unlikely 
high pattern-speed ($\Omega_p>$180\\kms$^{-1}$kpc$^{-1}$) which also implies a {\it marginal} ILR at 
r$\sim$4$\arcsec$. The latter is not compatible with the observations: the maximum distortion of the peanut 
bulge in NGC$\,4013$ is at r$\sim$12$\arcsec$. Moreover, as we ignore a priori the orientation of the bar major 
axis along the line of sight, r$\sim$12$\arcsec$ should be taken as a lower limit for the ILR.

Most of the available numerical simulations have treated the appearance of the peanut distortion and
its evolution including only the stars. Although the influence of a dissipative component
on the fate of the peanut remains to be studied thoroughly, a plausible
evolutionary sequence can be advanced here. The peanut instability first sets in at a {\it marginal} ILR of 
the bar, but this should be taken as a starting 
point. The subsequent inwards flux of gas towards 
the original ILR, driven by the bar's gravitational torque, can change the $\Omega$-$\kappa$/2 curve
in the inner region. The curve can become progressively steeper and finally two inner Linblad 
resonances (outer (oILR) and inner (iILR)) are bound to appear. The stellar bar itself is expected to 
slightly slow down in the process. The peanut distortion is the relic of an old {\it marginal}  
ILR; however a {\it strong} ILR region will develop in the course of time.
 Any reasonable value of $\Omega_p$ in NGC\,4013 implies we have two ILRs at present: 
an inspection of fig.5 leads to that conclusion unavoidably.
Therefore, the existence of a peanut instability and a {\it strong} ILR region in the nucleus may not 
be in contradiction, but should be taken as a result of secular evolution. 
Numerical simulations to be developed in paper II will allow us to test this scenario .    
       
Fitting the bar pattern speed is beyond the scope of the present paper,
 however a lower limit on $\Omega_p$ can be tentatively set. 
The radial distribution of molecular gas (with a relative depression or hole between r=10$\arcsec$ 
and r=30$\arcsec$) suggest that corotation cannot be 
at a radius larger than r$\sim$50$\arcsec$, implying that $\Omega_p>$60-70kms$^{-1}$kpc$^{-1}$.  
The latter implies the following loci of principal resonances: r$_{iILR}>$2$\arcsec$, 
r$_{oILR}<$12$\arcsec$, r$_{COR}<$50$\arcsec$, r$_{UHR}<$70$\arcsec$ and r$_{OLR}<$100$\arcsec$. 
Corotation of the bar is kept well inside the optical disk. This is expected 
for bars in spirals of early or intermediate Hubble types such as NGC\,4013, classified as Sbc 
(see Combes and Elmegreen 1993). Secondly, if $\Omega_p>$60-70kms$^{-1}$kpc$^{-1}$, the oILR appears 
inside the old marginal ILR (at 12$\arcsec$), as proposed lines above.   
Finally the major axis distribution of CO showing a maximum at r$\sim$50$\arcsec$ is well accounted for
if a ring forms near the UHR resonance (r$_{UHR}<$70$\arcsec$).

\section{A link between the gas filaments and the bar?}

 In a barred and peanut-shaped potential, the gas in principle tends to
follow the periodic orbits, and is dragged through vertical resonances
out of the plane, like the stars. However, the gas is dissipative,
on a short time-scale; cloud-cloud collisions occur with a characteristic
collision time of 10$^7$ yr. While the peanut and perpendicular
elevation time scale is a long process, of time-scale almost 10$^9$ yr
({\bf C90}). Therefore, the gas component looses its disordered
kinetic energy, and in particular in the z-direction, rather quickly.
The gas settles down in a very thin disk, irrespective of the peanut-shape
potential. 

In order to study if there can be a stable and spontaneous vertical response of gas 
in a peanut potential, we have performed self-consistent numerical simulations involving 
gas and stars. In this first run, the gas mass fraction is kept very low on purpose ($\sim$2\%). 
The gas clouds are considered as sticky particles and we neglected the effects of 
self-gravitation. No effect of star formation on the gas dynamics is considered either.
  The code used is FFT-based particle-mesh, with the gas treated as 
sticky particles (Combes \& Gerin 1985). The useful grid is 128$\times$128$\times$64,
and the suppression of the Fourier images is done through the algorithm of James (1977).
The run included 150k stellar particles and 40k gas particles. 
A bar forms spontaneously in the stellar component, 
since the bulge to disk mass ratio is 1/3, and there is no other spheroidal component 
that could stabilize the disk. After 2$\times$10$^9$ yr, the bar has developped a 
characteristic peanut shape (see contours of Fig.7).
The main result is that, at any time, the gas disk
remains very thin ($\sim$ 200pc at half-power), and is never
perturbed by the peanut/shaped potential or only in a very transient 
manner (see Fig. 8).

The vertical thickening of the bar and the subsequent formation of the peanut 
can be caused by instabilities associated with resonances between the bar 
motion and the vertical oscillations of the stars({\bf C90}; see also Binney(1981)); 
this effect has also been 
attributed to the buckling or fire-house instability by Raha et al 1991. 
The bar instability first acts to increase the eccentricity
of stellar orbits and align their principal axes; this causes the
buckling instability, precisely about the vertical resonance region,
which increases vertical velocity dispersion and thickness. Once
the bar is thickened, the nature of the stellar orbits in the peanut
can be described with orbits trapped around the 2:2:1 
periodic orbit family: these correspond to the vertical Lindblad
resonance, where in the frame of the bar, the particles just perform 
two z-oscillations in one turn ($\Omega - \nu_z/2 = \Omega_b$, 
the bar pattern speed).
This happens to occur in the region of the in-plane inner Lindblad
resonance (where $\Omega - \kappa/2 = \Omega_b$), and therefore
the resonant orbits have the 3-D shape of a banana (projected into
an ellipse in the plane ({\bf C90}, Pfenniger \& Friedli 1991).
A detailed orbital study has emphasized the bifurcations of the main
periodic orbits $x_1$ onto the banana and anti-banana orbit families
(Pfenniger \& Friedli 1991). These periodic orbits, in a strong bar,
very often possess loops, and the gas is not likely to follow them
because of dissipation. This may explain the small propensity of the gas
to be elevated vertically, at least for directly rotating orbits.
 For retrograde orbits, on the contrary, the orbit family $x_4$ bifurcates
due to vertical instability into the anomalous orbits (corresponding
this time to a 1/1 resonance, or one vertical oscillation
for one turn, Pfenniger \& Friedli 1991). This family does not possess
loops, and the gas can be maintained more easily on these (Friedli
\& Benz 1993). These retrograde orbits could be populated in particular
during an accretion event, where gas clouds arrive with a given 
angular momentum, un-correlated with that of the galaxy. 
The case of NGC\,128 is a remarkable example. It is a peanut-shaped
galaxy seen edge-on, where an inclined gaseous disk
is observed to counter-rotate with respect to the stars
(Emsellem \& Arsenault 1997). Although this galaxy has apparently
no sign of perturbed morphology, it has a companion nearby (NGC\,127)
which could have provided the retrograde gas.

\begin{figure}
\vspace{5mm}
\caption{Contours of stellar projected density of the bar seen edge-on,
after 4 Gyr of simulation. The scale is in kpc.}
\end{figure}

These stable retrograde orbits are characteristic of a tumbling
triaxial potential. Already van Albada et al (1982) had remarked
that in the rotating frame of a triaxial object, the Coriolis
forces on retrograde particles produce a torque, 
which is opposite to and compensates the torque of the restoring forces
towards the plane. The latter would have made the retrograde
orbits to precess, and since they differentially precess with 
radius, the clouds' collisions and dissipation would
have forced the gas to settle in the main plane. The Coriolis forces,
therefore, stabilise the gas into the inclined retrograde orbits.
 These orbits are not particularly related to the peanut shape, but 
both phenomena indicate the presence of bars.

In NGC\,4013, the presence of a prodigious warp (Bottema et al 1987)
suggests that a large amount of gas has recently
been accreted with un-related angular momentum.
Part of it could have gone towards the center. 
If spontaneously, directly-rotating gas will not stay at large height
above the plane due to the peanut shape of the potential, it is possible
that accreted gas with a different angular momentum, not aligned
with the principal axis, could be trapped in banana orbits, or more
likely in retrograde anomalous orbits. More simulations are needed,
taking into account the accretion origin of the gas, and also its larger mass fraction 
and self-gravity (see paper II).

An alternative mechanism to explain the presence of gas at high altitude
above the plane, is the galactic fountain effect, due to massive
star formation. This could explain the morphology of the H$\alpha$
filaments, that appear much higher above the plane than the neutral
gas, traced by the CO emission in the center. The fact that the
filaments seem to coincide with the peanut region might not
be a coincidence: the gas is easier to expel when the restoring force
of the plane is less, i.e. in the thicker stellar plane.

Extraplanar neutral+ionized gas and dust have been found in another edge-on spiral: NGC\,891.        
The existence of a thick molecular disk, containing M(H$_2$)$\sim$10$^{8}$M$\sun$ up to z$\sim$1kpc,
was first established by Garc\'{\i}a-Burillo et al (1992) using single dish CO observations. 
{\bf R96} also reports the detection of an extended DIG halo with some prominent H$\alpha$ filaments 
in this galaxy. Recently Howk and Savage (1997) have discovered the dusty counterpart of the thick 
CO disk: the HST and WIYN optical BVR images of NGC\,891 show hundreds of dust filaments lying 
far from the mid-plane. The derived neutral gas mass content of these absorbing structures 
closely agrees with the CO-based estimates. Although some of the extraplanar dust features are
interpreted as supernova-driven galactic chimneys, other are less clearly linked with highly energetic
phenomena in the disk. In either cases, the existence of an extended neutral+ionized gas with a set
of dust filaments is related to the high star formation efficiency in the disk 
(Garc\'{\i}a-Burillo et al 1992; {\bf R96}; Howk and Savage 1997).


\begin{figure*}
\vspace{5mm}
\caption{Particle plots at the same epoch (4 Gyr) of the stars (left)
and gas (right). Up is the face-on view, and bottom the edge-on view. In
each frame, the maximum radius is 25kpc.}
\end{figure*}
%
\section{Summary and Conclusions}

We have observed with the IRAM interferometer the emission of the 1--0 and 2--1 lines
of $^{12}$CO in the nucleus of NGC\,4013, an edge-on Sbc spiral possessing a box-shaped 
bulge, with spatial resolutions of $\sim$3.3$\arcsec$ and $\sim$1.2$\arcsec$, respectively.

Our maps show the presence of a distinct fast-rotating 
(v$_{rot}$\\$\sim$130kms$^{-1}$) nuclear 
disk of radius r$\sim$2$\arcsec$\,(100pc) and gas mass 
M$_{gas}\sim$0.6$\times$10$^{8}$M$_{\sun}$. The high-velocity component (absent in the HI map) 
is accompanied by gas emission at non-circular velocities within $\pm$3$\arcsec$ from the 
dynamical center, indicating that the gas flow is driven by a non-axisymmetrical potential. 
An analysis of the gas kinematics, derived from the $^{12}$CO(1--0) major axis position-velocity 
plot and the $^{12}$CO(2--1) isovelocities map, supports a bar model for NGC\,4013. 
The observed ring-like distribution of molecular gas (at r$\sim$2$\arcsec$\,(100pc) and 
r$\sim$50$\arcsec$\,(2.7kpc); the outer ring inferred from the 30m map) is interpreted as the 
imprint of a bar.

Although a link between the bar and the box-shaped bulge in NGC\,4013 is suggested,
there are apparent discrepancies between the results of numerical simulations and the model
proposed here. Disagreement concerns the basic parameters of the bar 
generating the peanut. In self-consistent simulations of the star component the peanut forms 
near the locus of a {\it marginal} ILR of the bar. Instead, the derived $\Omega$-$\kappa$/2 
curve in NGC\,4013 leaves little doubt of the existence of a {\it strong} ILR response, 
irrespective of the chosen pattern speed of the bar ($\Omega_p$). 
The nuclear disk would be located in the vicinity of the iILR, with no gaseous ring counterpart 
in the oILR. The present observations suggest that the inclusion of a dissipative component 
in simulations might probably change the evolution of the stellar peanut: 
although the latter appears near a marginal ILR, the inflow of gas driven by the bar, makes 
two ILRs appear and accelerates the secular evolution of the gaseous disk in the 
inner $\sim$1kpc region of the galaxy. Numerical simulations, involving
both the stars and the gas in a peanut potential, will analyse the response of the gas
in the disk of the galaxy fully testing the bar hypothesis (see paper II).  
      
We analysed the vertical distribution of molecular gas, showing that, although the 
bulk of CO emission comes from a thin disk (with deconvolved thickness 
FWHM$\sim$80-130pc), there is a non-negligible amount of molecular
 gas (M(H$_2$)=1.5$\times$10$^7$M$_{\sun}$) at large
z distances from the plane (z$\sim$200-300pc). The close relationship 
between the DIG filaments seen in H$\alpha$ coming out of the plane and the presence of 
molecular gas emission, suggests that both share a common origin: gas ejected by a massive 
nuclear starburst. 

A preliminary run of simulations has restricted to study the vertical response
of the gas to a peanut potential, that spontaneously forms in a disk of stars.
Gas clouds are treated as test particles and we neglect here the effects of star formation
and self-gravity in the gas dynamics. 
Due to its dissipative nature, the gas forms a thin disk very quickly. Contrary to the stars,
the gas cannot be maintained at high altitude above the galaxy plane along
stable orbits. Cloud-cloud collisions make impossible the population of banana and 
antibabana self-intersecting orbits. Moreover, it remains to be studied if gas clouds can
 populate the vertical bifurcation of the retrograde x$_4$ family in a peanut, after an accretion 
episode (see paper II). However this mechanism is unlikely to hold for NGC\,4013,
where the bulk of the gas in the disk is directly rotating.

Among the different explanations for the gas at high z--inclined resonant orbits
connected to the peanut, gas accretion in the course of an interaction and, finally, the galactic fountain model-- 
the latter seems the best to account for the H$\alpha$ and CO filaments. 
Although the peanut distortion formed in the stars 
comes from a bar in the disk (the presence of the latter being suggested by the observed CO kinematics), 
gas is being ejected in the nucleus after a bar driven starburst. The filaments come from the inner 
200pc(4$\arcsec$)
and reach a height of several Kpc, coinciding with the maximum peanut distortion where the strength of the
restoring forces of the plane is diminished.
  
\acknowledgements

This work has been partially supported by the Spanish CICYT under grant number 
PB96-0104. The authors heartly thank Richard Rand for giving us the H$\alpha$
image of NGC\,4013 used in this paper. We also would like to thank 
Martin Shaw for making available his K-band image. We finally thank James Binney,
the referee, for his encouraging comments and criticisms.

\end{document}